\documentclass[prd,nofootinbib,superscriptaddress,onecolumn,preprintnumbers]{revtex4}

\usepackage{amsmath}
\usepackage{amssymb}
\usepackage{float}
\usepackage[T1]{fontenc}
\usepackage{graphicx}
\usepackage{hyperref}

\usepackage{booktabs}
\usepackage{multirow}

\usepackage{color}

\begin{document}
\title{Viscous matter in modified $f(R)$-gravity:\\
Hubble tension attenuation with DESI and SH0ES data}

\author{Elisa Fazzari}
\email{elisa.fazzari@uniroma1.it}
\affiliation{Physics Department, ``Sapienza'' University of Rome,  P.le Aldo Moro 5, 00185 Roma, Italy}
\affiliation{Istituto Nazionale di Fisica Nucleare (INFN), Sezione di Roma, P.le A. Moro 5, I-00185, Roma, Italy}
\affiliation{Physics Department, Tor Vergata University of Rome, Via della Ricerca Scientifica 1, 00133 Roma, Italy}

\author{Nakia Carlevaro}
\email{nakia.carlevaro@enea.it}
\affiliation{Nuclear Department, ENEA - C. R. Frascati, Via E. Fermi 45, 00044 Frascati, Italy}

\author{Giovanni Montani}
\email{giovanni.montani@enea.it}
\affiliation{Nuclear Department, ENEA - C. R. Frascati, Via E. Fermi 45, 00044 Frascati, Italy}
\affiliation{Physics Department, ``Sapienza'' University of Rome,  P.le Aldo Moro 5, 00185 Roma, Italy}

\begin{abstract}
We study the late-time dynamics of the Universe in a modified metric $f(R)$ gravity model formulated in the Jordan frame. The standard $\Lambda$CDM model is additionally deformed to include a non-equilibrium effect on matter dynamics, represented via a bulk viscosity coefficient.
We impose an additional condition on the cosmological dynamics that allows the matter energy density, the non-minimally coupled scalar field, and its non-constant potential to be treated as dynamical variables. In this way, the form of the $f(R)$ model is not specified \emph{a priori}, but is instead dynamically reconstructed from the evolution of the system.
The resulting model contains five free parameters and is tested against Type Ia supernova and DESI data, using the SH0ES calibration for the supernova sample and a Planck calibration for the sound horizon. We find that the data mainly constrain the standard cosmological parameters $H_0$ and $\Omega_{m0}$, while the three additional parameters remain only weakly constrained and are further restricted by imposing the physical viability conditions of the model.
Our main result is that the model can accommodate SN Ia and DESI observations while allowing a larger value of $H_0$ than in the corresponding $\Lambda$CDM analysis and keeping the acoustic horizon close to the Planck value.
\end{abstract}

\maketitle

\section{Introduction}

In recent years, several observational results have challenged the ability of the $\Lambda$ Cold Dark Matter ($\Lambda$CDM) model \cite{weinberg-grav-cosm} to fully describe the late-time dynamics of the Universe. Two main observational results have particularly motivated the search for extensions of the standard cosmological scenario. The first is the more than $5\sigma$ discrepancy between the measurements of the Hubble constant $H_0$ obtained by the SH0ES \cite{SH0ES} and Planck \cite{Planck2018} Collaborations. The second comes from the DESI Collaboration \cite{desi,desi2} (see also \cite{nunes1,divalentino_interacting,colgain_Dainotti}), whose observations show a preference for a time-dependent dark energy equation of state described by the Chevallier--Polarski--Linder (CPL) parametrization \cite{CPL1,CPL2}, also known as the $w_0w_a$CDM model.

In particular, the latest DESI Data Release 2 (DESI-DR2) results \cite{desi2,desi_extended} provide new information on both the Hubble tension and possible departures from $\Lambda$CDM. Within the standard cosmological model, DESI-DR2 remains consistent with the Cosmic Microwave Background (CMB) angular scale, while showing a moderate tension with CMB constraints in the $(r_dH_0,\Omega_{m0})$ plane. Moreover, the value of $H_0$ inferred from DESI combined with Big Bang Nucleosynthesis (BBN) priors remains in strong tension with the SH0ES measurement, while differences also appear in the preferred values of the matter density parameter when DESI, Planck, and supernova data are compared. 

When the analysis is extended to the CPL parametrization, some of these internal tensions are reduced, and a dynamical dark-energy model is statistically preferred over the $\Lambda$CDM for several combinations of DESI, CMB, and Type Ia supernova (SN Ia) data \cite{desi2}. More generally, Ref.~\cite{Giare_Robust_DDE} showed that an evolving dark-energy scenario with phantom crossing is preferred when DESI data are included, independently of the specific parametrization adopted. This picture is also supported by model-independent cosmographic analyses \cite{Fazzari:2025lzd} and by non-parametric reconstructions of the dark-energy equation of state \cite{desi_extended,Gonzalez-Fuentes:2025lei,Nesseris:2012tt,Berti:2025phi} (see also Refs.~\cite{OColgain2025DES5YR,Teng2021DarkEnergyEoS}).

Although the CPL parametrization is favored over $\Lambda$CDM model by several current data combinations and may also help to alleviate the neutrino-mass tension that can arise in standard analyses \cite{Giare:2025ath}, it does not solve the Hubble tension and can even worsen it \cite{Linder:2024rdj}. Modified-gravity scenarios have therefore also attracted renewed interest (see Refs.\cite{Sotiriou_rev, Nojiri-odintsov, Faraoni:2010beyond} for general reviews about $f(R)$-modified gravity). For example, Ref.~\cite{DOnofrio:2025cuk} considered a framework including modified gravity and massive neutrinos, showing that it can remain consistent with present observations while providing an improved fit with respect to $\Lambda$CDM model. 

These results provide further motivation to investigate alternative cosmological models that can reproduce the observed late-time expansion history while remaining consistent with the full set of available observations. In this context, several different approaches have been explored, including dynamical dark energy models \cite{montani-carlevaro-dainotti2025,Kessler:2025kju, giare_overviewDDE, Navone:2025gxr, Schiavone:2026agq}, interacting dark matter-dark energy scenarios \cite{Montani:2024pou,Giare_interacting,MONTANI2025101969,divalentino_interacting}, and modified-gravity cosmologies \cite{schiavone2023,montani_hubbletension_5,matcre_montanimary,Nojiri-odintsov, DeLeo:2026luv} (see also Refs.~\cite{DymnikovaKhlopov1998,DymnikovaKhlopov2001,Mukhopadhyay--1,Doroshkevich-1,DoroshkevichKlypinKhlopov1989}). For an extensive review of late-time solutions to the Hubble tension, see Ref.~\cite{divalentino-Hubbletension}, while broader discussions of cosmological tensions, including early dark-energy scenarios, can be found in Refs.~ \cite{whitepaper_cosmoverse,review_early,odintsov3,DiValentino:2026uua,Krishnan2020EarlyUniverse}. 

Further indications that the standard $\Lambda$CDM description may be incomplete at late times have also emerged from binned analyses of SN Ia samples. These studies have reported an effective redshift dependence of the inferred Hubble constant \cite{dainotti2021,dainotti2022,Dainotti:2025qxz,greci_pantheon,Fazzari:2025mww,Schiavone:2024heb, Montani:2026bzk, carlevaro-letter}; similar conclusions have also been obtained using different cosmological probes \cite{dainotti2024binquasars,cinesi_H0z}. In particular, these analyses suggest a phenomenological power-law decrease of the effective value of $H_0$ with redshift. Such a behavior could originate from astrophysical evolution or systematic effects associated with the adopted sources, but it has also been interpreted in terms of modified gravity \cite{schiavone2023} and dark energy-dark matter interaction paradigm \cite{montani-entropy} (for a general reinterpretation of scaled $\Lambda$CDM model in terms of evolutionary dark energy see \cite{carlevaro-letter}).

A specific realization of metric $f(R)$ gravity was developed in Refs.~\cite{montani_hubbletension_5,matcre_montanimary, Fazzari:2025nfr,MONTANI2025101969} (see also Refs.~\cite{odintsov1,odintsov2} for general discussions of $f(R)$ cosmology). In this approach, the form of the gravitational Lagrangian is not assumed \emph{a priori}, but is reconstructed dynamically. More precisely, the potential of the non-minimally coupled scalar field emerging in the Jordan-frame representation \cite{Sotiriou,Sotiriou_rev,nojiri2011} is treated as an additional dynamical quantity, whose evolution is fixed by imposing an extra condition on the cosmological dynamics.

In this work, we develop an extension of the $\Lambda$CDM model aimed at describing both the phenomenology associated with the Hubble tension and the expansion history preferred by DESI observations. To this purpose, we combine two different ingredients. First, we consider a metric $f(R)$ modification of gravity in the Jordan frame. Second, we introduce a non-equilibrium contribution in the matter sector, described through an effective bulk-viscosity coefficient \cite{Belinskii1975,Belinskii1977,Belinskii1979,disconzi, carlevaro-letter,venanzi_bulk}. The combination of these two effects plays an important role in the consistency of the model. The bulk-viscous contribution alone does not provide enough freedom to reproduce the desired evolution of the Hubble parameter. On the other hand, the modified-gravity sector, when considered alone in the present formulation, can develop a tachyonic mode \cite{montani_hubbletension_5}. The simultaneous presence of modified gravity and non-equilibrium matter dynamics therefore allows us to construct a viable cosmological evolution while introducing enough flexibility to reproduce the late-time phenomenology suggested by current observations. In this sense, the model combines a fundamental modification of the gravitational sector with an effective description of non-equilibrium processes in the cosmic matter component. We test the model against late-time observations through a Markov Chain Monte Carlo analysis, combining the Pantheon+ SN Ia sample calibrated with SH0ES Cepheids and the DESI-DR2 BAO measurements anchored to the Planck sound horizon. The data leave the additional parameters of the model only weakly constrained; we therefore fix them by imposing, at a background level, theoretical viability conditions, namely the absence of tachyonic modes and a matter density profile compatible with standard structure formation. For the same data combination, the model yields $H_0 = 69.62 \pm 0.52\,\mathrm{km/s/Mpc}$, compared with $H_0 = 65.8 \pm 1.6\,\mathrm{km/s/Mpc}$ in a $\Lambda$CDM scenario, while keeping the sound horizon close to its Planck value. The Hubble tension is thus attenuated through a reshaping of the late-time expansion history, without modifying pre-recombination physics. The reconstructed $f(R)$ Lagrangian remains a weak deformation of the Einstein-Hilbert action plus a cosmological constant.

The manuscript is organized as follows. In Section~\ref{sec2}, we introduce the modified cosmological framework and derive its main dynamical equations. In Section~\ref{sec3}, we present the data analysis part and finally, our conclusions are reported in Section~\ref{sec4}.

\section{Modified cosmological dynamics}\label{sec2}
We examine a metric $f(R)$ theory formulated in the Jordan frame, where deviations from General Relativity (GR) are encapsulated by a scalar field $\phi$ with non minimal coupling. The action for this theory is given by
\begin{equation}
    S = \frac{1}{2\chi}\int d^4x 
    \sqrt{-g}\left( \phi R - V + 2\chi \mathcal{L}_m\right)\;,
    \label{eq: action}
\end{equation}
where $\chi$ denotes the Einstein constant, while $g$ and $R$ represent the determinant and Ricci scalar of the metric tensor $g_{\mu\nu}$, respectively, and $\mathcal{L}_m$ stands for the matter Lagrangian density. The potential $V(\phi)$ is determined by the particular $f(R)$ function through the relation
\begin{equation}
   \phi= f_R(\phi), \quad V(\phi ) \equiv \phi R(\phi )- f(R(\phi )), \quad  \quad R(\phi ) = f_R ^{-1}(\phi)\;,
    \label{rt2}
\end{equation}
where $f_R=df/dR$ must be invertible. Importantly, $g_{\mu\nu}$ and $\phi$ constitute independent degrees of freedom, and the action (\ref{eq: action}) must be varied with respect to both of them, in addition to matter fields.

We now apply this modified gravity framework to the late Universe, neglecting spatial curvature and radiation contributions \cite{efstathiou_planck}. The geometry is described by a Robertson-Walker metric \cite{weinberg-grav-cosm}
\begin{equation}
    ds^2 = -dt^2 + a(t)^2\left( dx^2 + dy^2 + dz^2\right),
    \label{rt3}
\end{equation}
where Cartesian coordinates are used, and the cosmic scale factor $a$ evolves with synchronous time $t$. According to the analysis in \cite{montani_hubbletension_5}, we consider
\begin{align}
V(\phi) = 2\chi \rho_{\Lambda} + G(\phi)\;,   
\end{align}
where $\rho_{\Lambda}$ the present day vacuum energy density and $G$ a generic function. Denoting with $H(t)\equiv a^{-1}da/dt$ the Hubble parameter, it is convenient to use the redshift variable $z$ defined as $a_0/a=1+z$, where here and in the following the suffix $0$ denotes the present day ($t=0$) value of a quantity. Furthermore, the 00 component of the field equations known as the Friedmann equation is split into two different components in order to remain as close as possible to $\Lambda$CDM dynamics. This splitting also allows usto treat the function $G(\phi(z))\equiv G(z)$ as a dynamical variable. By varying the action with respect to $\phi$, we get the relation $R=dG/d\phi$ which, implemented in the considered cosmological setting, yields the third dynamical equation. Hence, the system dynamics is provided by the following equations
\begin{align}
H^2 &= \frac{\chi}{3\phi}\left( \rho_m + \rho_{\Lambda}\right)\;,\label{ie1}\\
\phi ' &= -\frac{G}{6H^2(1+z)}\;,\label{ie3}\\
\frac{dG}{d\phi}& = 12 H^2 - 6(1+z)HH'\;,\label{ie2}
\end{align}
where the prime denotes differentiation with respect to $z$ and $\rho_m$ stands for the matter energy density. Here, we assume that the matter contribution is subject to non-equilibrium effects, described via a bulk viscosity coefficient $\xi(\rho_{m})$ within the hydrodynamical approach. Bulk viscosity modifies the energy-momentum tensor to account for non-ideal effects in a compressible fluid. When a fluid expands or contracts, this term introduces an effective pressure $\Pi=-\xi \nabla_\mu u^\mu$, which is negative in an expanding Universe meaning that it that it resists to those changes. The fluid resists expansion and compression, generating a dissipative effect that tends to restore the system to equilibrium, i.e. the energy momentum tensor reads
\begin{equation}
T_{\mu \nu} = (\rho_m + P_m + \Pi)u_\mu u_\nu + (P_m+\Pi) g_{\mu \nu} \;,
\end{equation}
where $P_m$ is the standard thermodynamical pressure and it can be neglected for a matter contribution, while $\Pi=-\xi \nabla_\sigma u^\sigma$ ($u^\sigma=(1,0,0,0)$ is the comoving fluid four velocity). Thus, the matter dynamics is governed by the following equation \cite{Primordial}:
\begin{equation}
\rho_{m}' = 3\frac{\rho_{m} -	3H\xi}{1+z}\;,\label{ie4}
\end{equation}
and, in this work, we use the expression $\xi=\bar{\xi}\rho_m^\nu$, where $\bar{\xi}$ is a constant and, to have a perturbative bulk viscosity contribution, we assume $0\leq \nu\leq 1/2$ \cite{Belinskii1975,Belinskii1979}.

We now introduce the dimensionless quantities
\begin{equation}
E\equiv \frac{H}{H_0}\;,\qquad 
\Omega_m\equiv \frac{\chi\rho _m}{3H_0^2}\;,\qquad
\Omega_{\Lambda}\equiv \frac{\chi\rho_{\Lambda}}{3H_0^2}\;,\qquad
\Omega_G \equiv \frac{G}{6H_0^2}\;,\label{ie6}
\end{equation}
and the dynamics governing the late Universe expansion can be recast via the following dimensionless form:

\begin{align}
E^2 &= \phi^{-1}(\Omega_m(z) + \Omega_{\Lambda})\;, \label{ie7}\\
\phi' &= - E^{-2}\Omega_G (1+z)^{-1}\;, \label{ie8}\\
\Omega_G' &= \phi' \left(2 E^2 - E E' (1+z) \right)\;, \label{ie9}\\
\Omega_{m}' &= 3 \Omega_{m} \left( 1 -\bar{\Omega}^{1/2-\nu}\,E\, \Omega_{m}^{\nu - 1} \right)(1+z)^{-1}\;, \label{ie10}
\end{align}
where, we impose the present-day normalization $\phi_0=1$ and  $\Omega_\Lambda=1-\Omega_{m0}$, and $\bar{\Omega}=3^\nu\chi^{1-\nu}H_0^{2\nu-1}\,\bar{\xi}$.

By dividing Eq.(\ref{ie8}) with Eq.(\ref{ie9}), we obtain the solution for the dimensionless potential:
\begin{align}
\Omega_G(z)=-A^{2^*}\,E\,(1+z)^{-2}\,,
\label{eqie11}
\end{align}
with $A^{2^*}\equiv - \Omega_{G0}$.

Finally, the evolutive system can be written as follows 
\begin{align}
E&= \phi^{-1/2}(\Omega_m + \Omega_{\Lambda})^{1/2}\;, \label{ie12} \\
\phi'&=A^{2^*}\,E^{-1} (1+z)^{-3}\;, \label{ie13}\\
\Omega_{m}' &= 3 \Omega_{m} \left( 1 -\bar{\Omega}^{1/2-\nu}\,E\, \Omega_{m}^{\nu - 1} \right)(1+z)^{-1}\;.
\end{align}
We stress how the physical model we set has five free parameters: $H_0$, $\Omega_{m0}$, $\nu$, $A^{2^*}$ and $\bar{\Omega}$. It is worth noting that the system above does not explicitly contain the Hubble constant since we deal with the expansion rate $E$.

We conclude this section by mentioning that the model here proposed is viable only if it does not contain any tachyon modes, i.e. the scalar-field mass squared
\begin{align}
\mu^2_{\phi}\equiv \frac{1}{3}\left( \phi \frac{d^2V}{d\phi ^2} - \frac{dV}{d\phi}\right) \;,
\label{massss}
\end{align}
must remain positive.

\section{Data analysis}\label{sec3}
In this section, we compare the predictions of our model with observational data and impose the physical viability conditions discussed in the previous section. We first use a Markov Chain Monte Carlo (MCMC) analysis to constrain the model parameters. We then study the parameters that remain weakly constrained by the data by imposing the theoretical conditions required for a physically viable cosmological evolution.

\subsection{Dataset}
We test the modified model introduced above and constrain its free parameters, namely $H_0$, $\Omega_{m0}$, $\bar{\Omega}$, $A^{2^*}$, and $\nu$. Our main goal is to investigate whether the model can accommodate a larger value of the Hubble constant while providing a good description of the DESI observations.

For this purpose, we performed an MCMC analysis using the publicly available sampler \texttt{Cobaya} \cite{cobaya}, interfaced with our numerical code that computes the theoretical predictions of the modified model. We require the Gelman--Rubin convergence criterion $R-1<0.01$ \cite{gelman_rubin}. The statistical analysis of the chains and the posterior distributions are obtained using the \texttt{GetDist} package \cite{getdist}.

We adopt flat priors for the standard cosmological parameters, $H_0 \in [20,100] \mathrm{km/s/Mpc}$  and $\Omega_{m0}\in[0.01,0.99]$. For the additional parameters of the model, we choose $\bar{\Omega}\in[10^{-5},10^{-1}]$, since this quantity is expected to represent a fraction of the matter contribution, $A^{2^*}\in[0,1]$, because it directly affects the positivity of the scalar-field mass, and $\nu\in[0,0.5]$, motivated by the non-equilibrium effects discussed in Section~\ref{sec2}.

All the adopted prior ranges are summarized in Table~\ref{tab:prior}.

\begin{table}[h]
\centering
\begin{tabular}{ll}
\toprule
\textbf{Parameter} & \textbf{Prior} \\
\midrule
$\Omega_{m0}$ & $\mathcal{U}[0.01,\,0.99]$ \\
$H_0\,[\mathrm{km/s/Mpc}]$\;\;\;\;\; & $\mathcal{U}[20,\,100]$ \\
$log_{10}\bar{\Omega}$ & $\mathcal{U}[-5,\,-1]$ \\
$A^{2^*}$ & $\mathcal{U}[0,\,1]$ \\
$\nu$ & $\mathcal{U}[0,\,0.5]$ \\
$r_d \, [\mathrm{Mpc}]$ & $\mathcal{N}[147.09, 0.26]$ \\
\bottomrule
\end{tabular}
\caption{Prior ranges considered in the analysis.}
\label{tab:prior}
\end{table}

Since our main interest is the late-time expansion of the Universe, and in particular the Hubble tension and the DESI results, we consider the following datasets:

\begin{itemize}

\item \textit{Type Ia Supernovae with SH0ES calibration}: Distance moduli measurements from the PantheonPlus catalog \cite{Scolnic_2022, Brout_2022} with the SH0ES Cepheid host distances used to calibrate the SN Ia sample \cite{SH0ES}. The sample consists of 1701 light curves for 1550 uncalibrated SNe Ia spanning a redshift range of $0.01$ to $2.26$. We denote the SH0ES calibrated sample as ``SH0ES''. Data has been taken from the public repository \cite{SN_git}.
    
\item \textit{DESI-DR2 BAO}: Measurements of Baryon Acoustic Oscillations (BAO) from DESI's second data release \cite{desi2}, with effective redshifts of the BAO measurements ranging from $z_{\mathrm{eff}}=0.295$ to $z_{\mathrm{eff}}=2.33$. This sample was calibrated with Planck's acoustic horizon $r_d^{\mathrm {Planck}} = 147.09\pm0.26$ Mpc as reported in Table 2 of Ref.~\cite{Planck2018}. We refer to this dataset as ``DESI + $r_{\rm{d^{Planck}}}$''.

\end{itemize}

\subsection{Results}
\label{sec:results}
\begin{figure}[ht!]
    \centering
    \includegraphics[width=0.75\linewidth]{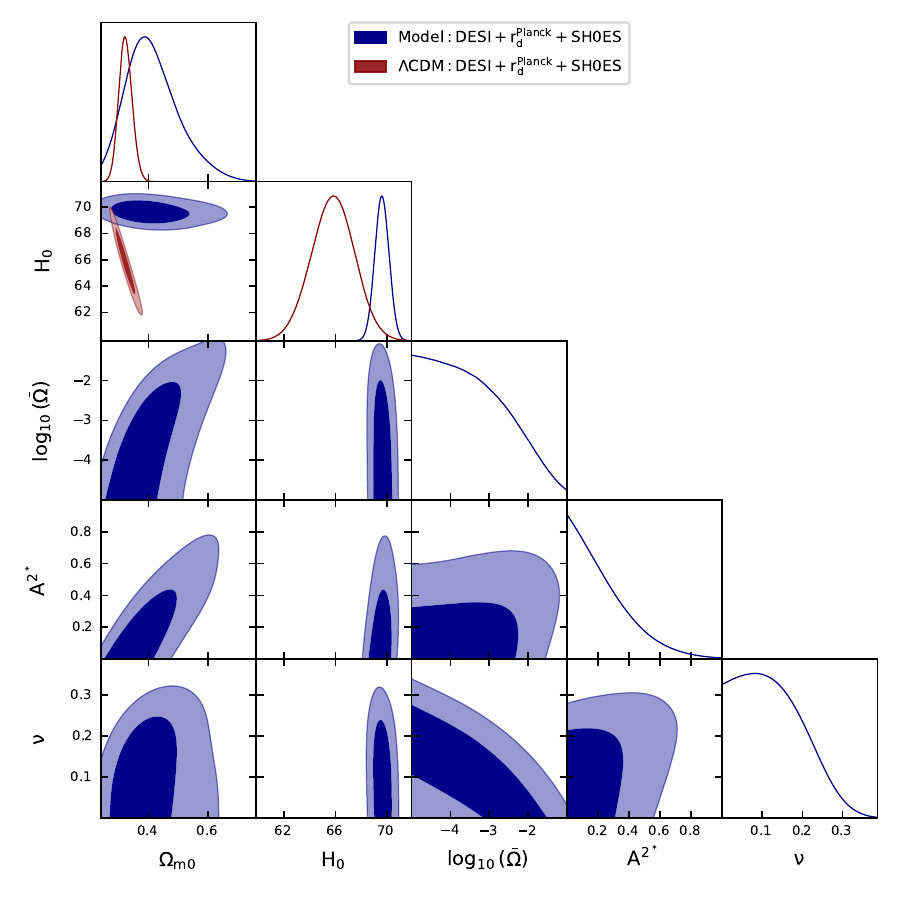}
    \caption{One-dimensional posterior probability distributions and two-dimensional $68 \%$ and $95\%$ CL contours for the free parameters of the modified model $H_0$, $\Omega_{m0}$, $A^{2^*}$, $\nu$ and $\log_{10}\bar\Omega$ and for the $\Lambda$CDM. Constraints are derived from a joint analysis combining PantheonPlus Type Ia supernovae calibrated with SH0ES and DESI data calibrated using the Planck 2018 results \cite{Planck2018}.}
    \label{fig:fR_bulk}
\end{figure}

We now discuss the constraints obtained from the statistical analysis described above. The marginalized constraints for the reference $\Lambda$CDM model and for the modified model are reported in Table~\ref{tab:Results_Combined}, while the corresponding one- and two-dimensional marginalized posterior distributions are shown in Fig.\ref{fig:fR_bulk}.

The posterior distributions shown in Fig.~\ref{fig:fR_bulk} together with the values reported in Tab.\ref{tab:Results_Combined} show that the extra parameters of the modified model are only weakly constrained by the dataset considered. At $95\%$ credibility, $A^{2^*}$ and $\nu$ are constrained only through upper limits. The parameter $\log_{10}(\bar{\Omega})$ is also bounded only from above. Moreover, the broad two-dimensional contours show significant degeneracies among the additional parameters. Therefore, the current data do not provide evidence for a non-zero value of any individual extra parameter.

Overall, the main phenomenological effect of the modified model is visible in the standard cosmological parameters. In particular, the additional freedom introduced by the modified gravitational and non-equilibrium sectors allows the model to accommodate a larger value of $H_0$ while keeping the sound horizon close to the value inferred by Planck. The relative uncertainty on $H_0$ also decreases from $2.43\%$ in $\Lambda$CDM to $0.75\%$ in the modified model, indicating a tighter constraint under the adopted model assumptions and priors.

We do not report an information criterion comparison, such as the Bayesian Information Criterion (BIC) \cite{wagenmakers2007}, between the modified model and $\Lambda$CDM since, in our analysis, several of the additional parameters are only weakly constrained by the MCMC data analysis. They are subsequently restricted mainly by the theoretical viability conditions discussed above, but this second step does not define an additional likelihood and therefore does not provide a corresponding value of $\chi^2$. As a consequence, a BIC computed only from the MCMC stage would not consistently describe the final physically viable parameter space considered in this work. We therefore restrict our discussion to the posterior constraints and to the physical viability of the model, without assigning a statistical preference between the modified model and $\Lambda$CDM through the BIC.

\begin{table}[h]
\centering
\begin{tabular}{lcccccc}
\toprule
\noalign{\vskip 0.15em}
 &  $\mathbf{H_0}\,  [\mathrm{km/s/Mpc}] $ & $\mathbf{\Omega_{m0}}$& $\mathbf{A^{2^*}}$& $\boldsymbol{\nu}$& $\mathbf{\log_{10}(\bar{\Omega})}$& $\mathbf{r_d} \, [\mathrm{Mpc}]$
\\[-0.1em]
\midrule
$\rm Model$ &\qquad $69.62\pm0.52$  &\qquad $0.410^{+0.071}_{-0.10}$ &\qquad $<0.600$&\qquad $<0.260$ & \qquad$<-1.82$ &\qquad $146.93\pm0.25$ 
\\ [0.4em]
$\Lambda$CDM &\qquad$65.8\pm1.6$&\qquad $0.322\pm0.021$&\qquad -- &\qquad -- &\qquad --&\qquad $146.18\pm0.24$\\

\noalign{\vskip 0.15em}
\hline
\end{tabular}

\caption{Marginalized constraints for the $\Lambda$CDM and the modified model obtained from the combination of PantheonPlus Type Ia supernovae calibrated with SH0ES and DESI data calibrated with the Planck sound horizon $r_d^{\rm Planck}$. Reported uncertainties on $H_0$, $\Omega_{m0}$, and $r_d$ correspond to $68\%$ marginalized credible intervals, while constraints on $A^{2^*}$, $\nu$, and $\log_{10}(\bar{\Omega})$ are quoted at $95\%$ C.L. . A dash indicates that the corresponding parameter is either not present in the model or not constrained.}
\label{tab:Results_Combined}
\end{table}

\newpage

\newpage

\subsection{Study of the parameter space}
\begin{figure}[ht!]
 \centering
    \includegraphics[width=0.49\linewidth]{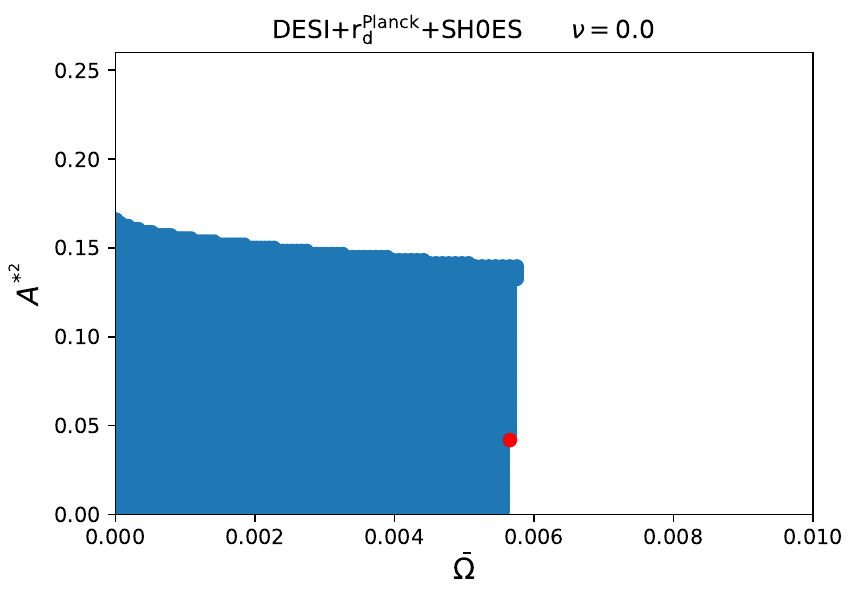}
    \includegraphics[width=0.475\linewidth]{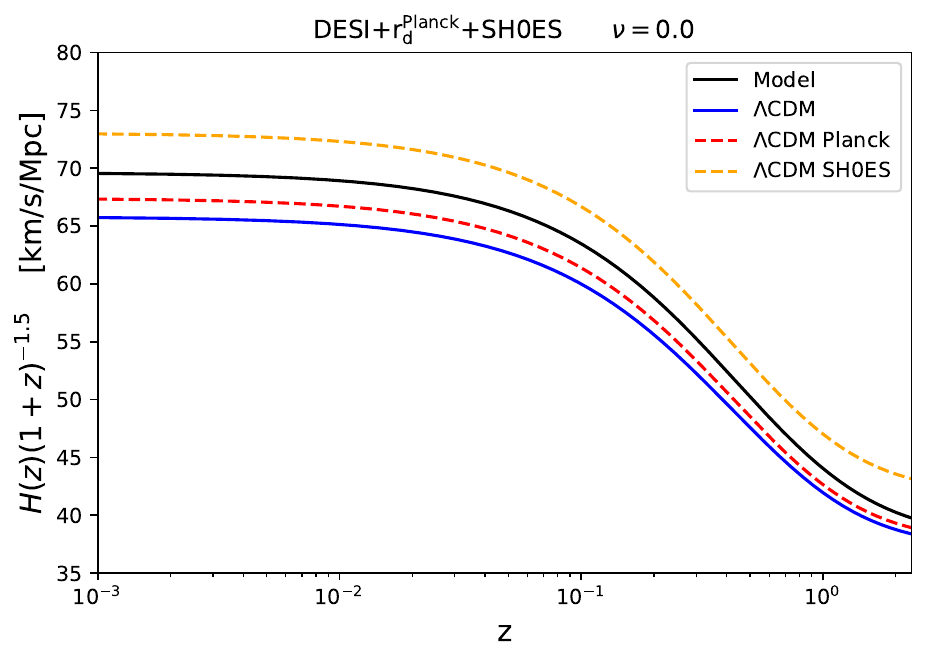}
    \caption{DESI+SH0ES+$r_d^{Planck}$ and $\nu=0$. Left-hand panel: parameter space $(\bar{\Omega},A^{2^*})$, the available domain is indicated in blue while the red dot corresponds to the analyzed case (we set $H_0=69.62$ and $\Omega_{m0}=0.41$). Right-hand panel: normalized Hubble parameters as indicated in the plot (details in the text).}
    \label{fig:figs_planck}
\end{figure}

Since the data analysis procedure outlined that the three parameters $\nu$, $\bar{\Omega}$ and $A^{2^*}$ are only weakly constrained, with the data providing upper limits rather than localized two-sided constraints. Therefore, we are legitimated to theoretically define the values of these parameters according to suitable prescriptions, by keeping set the values of $\Omega_{m0}$ and $H_0$ as the best fit in Tab.II. First of all, for the viscosity coefficient we make the simplest and, probably, most reasonable hypothesis of taking it as a constant, i.e. we fix $\nu=0$. For what concerns the parameter plane $(\bar{\Omega},A^{2^*})$, we restrict the viable domain by requiring the avoidance of tachyon modes, i.e. $\mu^2_\phi\ge0$. Furthermore, we impose that, at the simulation ending redshift $z^*=2.33$ (covering all the largest effective redshift of the BAO measurements used), the effective critical mass parameter $\Omega_m(z)/(1+z)^3$ is at least the Planck value $\Omega_{m0}^P=0.315$. This choice is physically motivated by the necessity to preserve as much as possible the standard profile of the matter density as required for a satisfactory structure formation process \cite{padmana87}. This background level viability procedure yields the domain sketched in the left-hand panel of Fig.\ref{fig:figs_planck}. 

In order to define a specific point in the obtained parameter space domain, we ask to minimize the value $\Omega_m(z)/(1+z)^3$: we thus arrive to determine the point corresponding to $(0.00565,0.04188)$. Setting $\nu=0$, we plot the associated (normalized) Hubble parameter in the right-hand panel of Fig.\ref{fig:figs_planck}, where we also overplot the $\Lambda$CDM curve inferred from the data in the previous subsection. For reference we also depict the standard flat $\Lambda$CDM model Hubble parameters constrained by the CMB data from Planck \cite{Planck2018}: $H_{0}^P= 67.4 \,\mathrm{km/s/Mpc}$, $\Omega_{m0}^P$ (red dashed), and by the Type Ia SN data calibrated with Cepheids from the SH0ES collaboration \cite{SH0ES}: $H_{0}^S=73.04 \,\mathrm{km/s/Mpc}$, $\Omega_{m0}^S= 0.331$ (orange dashed). 
\begin{figure}[ht!]
 \centering
    \includegraphics[width=0.48\linewidth]{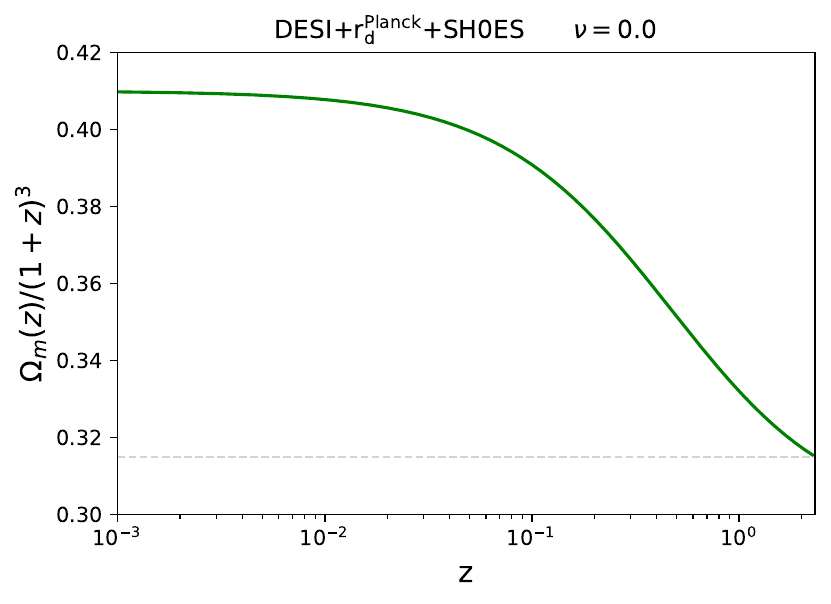}
    \includegraphics[width=0.49\linewidth]{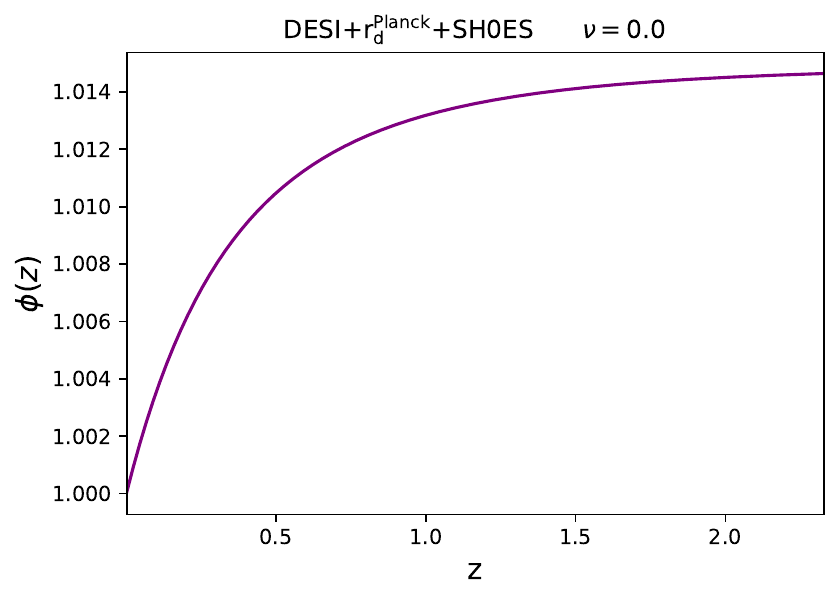}\\
    \includegraphics[width=0.49\linewidth]{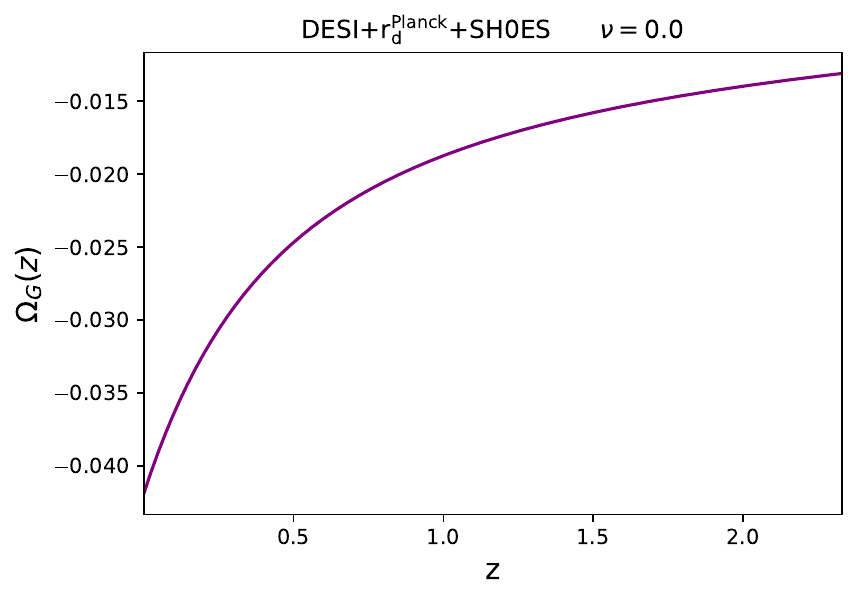}
    \includegraphics[width=0.465\linewidth]{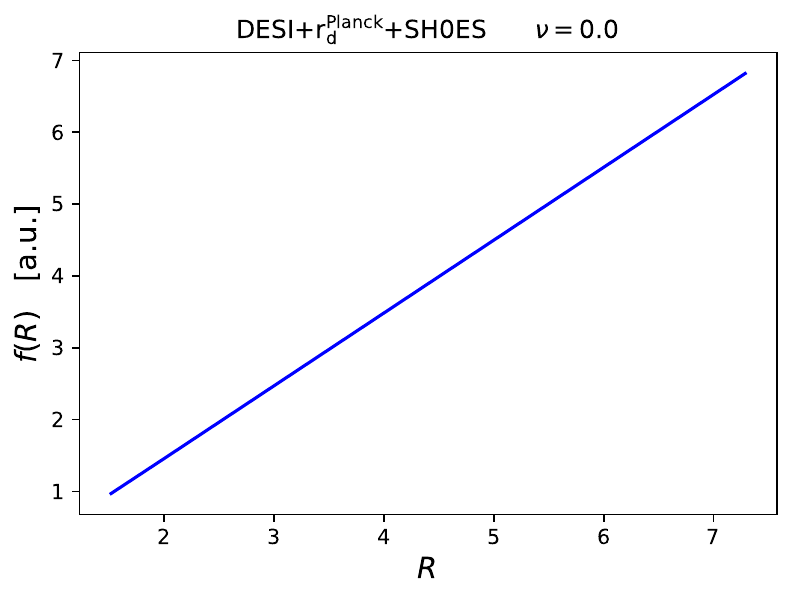}
    \caption{DESI+SH0ES+$r_d^{Planck}$ and $\nu=0$. Left-hand top panel: evolution of the (normalized) matter density parameter (the gray dashed line indicates the minimum). Right-hand top panel: plot of the scalar filed dynamics. Left-hand bottom panel: evolution of the normalized potential $\Omega_G(z)$. Right-hand bottom panel: plot of the $f(R)$ function in arbitrary units for the considered redshift domain.}
    \label{fig:figs_planck_}
\end{figure}

Within our model, the joint fit of the addressed sources yields a value of $H_0$ systematically higher than the one obtained from the same data combination under $\Lambda$CDM, thereby reducing the discrepancy with the local distance-ladder determination. It is important to note that, since the BAO standard ruler is anchored to the standard CMB-derived sound horizon, the model does not alleviate the tension by rescaling the pre-recombination physics ($r_{\mathrm{d}}$ is constrained by a Gaussian prior centered on the Planck value) but rather by reshaping the late-time expansion history, so that the SN Ia and BAO distances can be jointly accommodated with a larger $H_0$. The result should therefore be read as an effectively late-time attenuation of the tension: the combined constraint moves toward, but does not coincide with, the SH0ES value, leaving a residual offset on $H_0$. That such a shift is obtained without modifying the sound horizon is itself notable, given that purely late-time modifications of the expansion history are known to be tightly constrained by the shape of the low-redshift distance data.

Finally, in Fig.\ref{fig:figs_planck_}, we plot the evolution of $\Omega_m(z)$ indicating the corresponding minimum values, the scalar field $\phi(z)$, the potential term $\Omega _G(z)$ and the underlying $f(R)$ model. The scalar field outlines a monotonically increasing behavior, approaching an asymptotic value. This feature is just associated with the shift in the inferred Hubble constant, since it induces a corresponding common scaling of all the Universe energy density components. Also the potential term $\Omega_G(z)$ is characterized by a small global variation (we recall that it is the deviation of the modified gravity from a simple cosmological constant), approaching an asymptotic configuration. The physical consistence of the proposed modified $f(R)$ gravity picture relies on the very small deviation that the corresponding curves exhibit with respect to a straight line with an intercept of the order $1-\Omega_{m0}$. In fact, such a behavior ensures that, apart from the presence of a cosmological constant term, the modified gravity model corresponds to a very weak deformation with respect to an Einstein-Hilbert Lagrangian. These results provide a safe gravitational ground, in the sense that the proposed Lagrangian accounts for the cosmological constant term, accelerating the Universe, without introducing exotic effects in the physics of the gravitational interaction.

\section{Conclusions}\label{sec4}

We constructed an alternative paradigm to the $\Lambda$CDM model for the late Universe evolution, by combining two different physical effects: i) a modified metric $f(R)$ gravity in the Jordan frame (whose form was dynamically determined); ii) a non-equilibrium effect, which is altering the standard dynamics of the matter Universe component (represented via a bulk viscosity coefficient). We developed the detailed dynamics of the model via an additional condition, which allowed to treat the potential term of the non-minimally coupled scalar field as a dynamical quantity to be determined \emph{a posteriori} thereby prescribing the underlying modified gravity picture. Neither effect alone can provide the desired dynamics: the self-interacting scalar field mainly accounts for a rescaling of the Hubble parameter, while the bulk viscosity corresponds to a direct deformation of the standard ratio of the matter to the dark energy components.

The main result of our study is that, within the framework of our
cosmological model, the joint fit of the SH0ES-calibrated Pantheon+ sample and the DESI BAO data calibrated on the Planck  sound horizon yields $H_0 = 69.62 \,\mathrm{km/s/Mpc}$, to be compared with $H_0 = 65.8 \,\mathrm{km/s/Mpc}$ obtained from
the same data combination within standard $\Lambda$CDM model. The model
therefore drives the inferred expansion rate upward, shifting the combined
constraint toward the local distance-ladder value and thus attenuating
the Hubble tension. We stress that this attenuation is achieved without altering the pre-recombination physics: the sound horizon is constrained by a Gaussian prior centered on the Planck value, and the higher $H_0$ arises
entirely from a reshaping of the late-time expansion history that allows
the supernova and BAO distances to be jointly accommodated. The residual
offset with respect to the SH0ES determination shows that the tension is
mitigated rather than fully resolved, and a definitive assessment will
require folding in the full CMB likelihood and the growth-of-structure
constraints, which we leave to future work. Furthermore, we observe how the non-equilibrium expansion of the dark matter, as described by the bulk viscosity coefficient, is also responsible for the emergence of an effective parameter $\Omega_{m}(z)/(1+z)^3$. This effect should be carefully considered when studying the evolution of the large scale structure in the considered redshift interval, with a potential impact on the so-called $\sigma$8 tension.

These results should be regarded as a starting point in view of selecting those classes of new physics model, which are more promising to significantly attenuate, or even completely solve, the Hubble tension.

\section*{Acknowledgments}
E. F. acknowledges the IT Services at The University of Sheffield for providing High Performance Computing resources. We would like to thank Eleonora Di Valentino for her valuable suggestions on the model testing procedure.


\end{document}